# Reed-Muller Realization of *X (mod P)*

Danila A. Gorodecky
United Institute of Informatics Problems of
NAS of Belarus
Minsk, Belarus
danila.gorodecky@gmail.com

***Abstract*—This article provides a novel technique of *X (mod P)* realization. It is based on the Reed-Muller polynomial expansion. The advantage of the approach concludes in the capability to realize *X (mod P)* for an arbitrary *P*. The approach is competitive with the known realizations on the speed processing. Advantages and results of comparison with the known approaches for *X [9:1]* and *P=7* is demonstrated.**

## I. INTRODUCTION

The realization of the *X (mod P)* operation occupies a central place in cryptography; an efficiency of its realization in the residue number system (RNS) defines whether RNS will find wide implementation in practice or not.

There are two ways for hardware realization of *X (mod P)*. The pipelining realization is implemented for transformation of a data flow to a sequence of residues. An example of a fast pipelining realization in cryptography has been proposed in [1]. Another way is based on the iteration process. An iteration produces the bits in decreasing significance with later iterations producing bits of lower significance. Variations of these techniques referring to RNS have been proposed in [2, 3]. The main goal of the realization *X (mod P)* is to achieve high speed processing.

The first way is suitable for an arbitrary value of *P*, but the speed of the approach is limited by a block of pipeline, which includes three kind of successive operations (comparison, multiplexing, and subtraction). The second way is efficient just for some types of $P$ ($2^n, 2^n \pm 1, 2^n \pm 3$ [2,3,4] and some variations of them [5], e.g. $2^n \pm 2^{{n+1}/2} \pm 1, 2^{2n+1} \pm 1$). The article proposes an approach for *X (mod P)* realization which is suitable and efficient for an arbitrary value of *P*, but it competitive with an iterative procedure.

Result of the calculation of *X (mod P)=S* is $\delta$-bits binary vector, where $\delta = \lceil \log_2 P \rceil + 1$, and every digit of $S = (S_\delta, S_{\delta-1}, ..., S_1)$ is a Boolean function represented by Zhegalkin (or positive polarity Reed-Muller) form – polynomial XOR expansion with only incompetent variables or Zhegalkin expansion. The rest of the material describes a technique of generation of the polynomial extensions and hardware realizations of them.

A case for $P = 2^\delta$ does not considered due to a simple way of realization of $X(\mathrm{mod}\, 2^\delta)$. In this way $(x_\delta, x_{\delta-1}, ..., x_1)(\mathrm{mod}\, 2^\delta) = (x_n, x_{n-1}, ..., x_1)$, i.e. $X = (x_{10}, x_9, ..., x_1)$ and $P = 2^3$, then $(x_3, x_2, x_1)(\mathrm{mod}\, 8) = X$.

## II. X (MOD P) REALIZATION BY USING BOOLEAN FUNCTIONS

The idea of the approach is to consider a result of *X (mod P)=S* as the system of $\delta$ Boolean functions. Let's define $X = (x_n, x_{n-1}, ..., x_1)$, $P = (p_\delta, p_{\delta-1}, ..., p_1)$, and $S = (S_\delta, S_{\delta-1}, ..., S_1)$. Hence a Boolean function $S_i$, $i = \overline{1, \delta}$, depends on $n$ variables, i.e. $S_i = S_i(x_n, x_{n-1}, ..., x_1)$.

For any other case For example, a system of functions for the case *X (mod 5)=S*, where $X = (x_5, x_4, x_3, x_2, x_1)$, takes the following form:

$$\begin{cases} S_1 = S_1(x_5, x_4, x_3, x_2, x_1); \\ S_2 = S_2(x_5, x_4, x_3, x_2, x_1); \\ S_3 = S_3(x_5, x_4, x_3, x_2, x_1). \end{cases}$$

An arbitrary Boolean function depends on *n* variables may be represented with the set of the *truth numbers* $A(S_i)$ – numbers which correspond to the indexes of the truth table vector $w(S_i)$. This set contains numbers corresponded to unities on function values. Let's generate Boolean functions with the following way: a function $S_i(x_n, x_{n-1}, ..., x_1) = 1$ if and only if $X(\mathrm{mod}\, P) = S(S_\delta, S_{\delta-1}, ..., S_i = 1, ..., S_1)(\mathrm{mod}\, P)$. It is an equivalent for the set of the truth numbers $A(S_i)$, when this set consists of numbers contained unity on $i-$th bits on *mod P* in the range from 0 to $2^n - 1$. For example, for $X = (x_3, x_2, x_1)$ and $P = 3$, the set of truth numbers for $S_1$ is equal $A(S_1) = \{1,4,7\}$ and for $S_2$ is equal $A(S_2) = \{2,5\}$.

There is a one to one correspondence between the set of the truth numbers $A(S_i)$ and the truth vector $w(S_i)$: the $j$-th entry of the set of truth numbers corresponds to the $j$-th unity of the truth vector. Thus $A(S_1) = \{1,4,7\}$ is transformed into $w(S_1) = \{0,1,0,0,1,0,0,1\}$ and $A(S_2) = \{2,5\}$ is transformed into $w(S_2) = \{0,0,1,0,0,1,0,0\}$. Let's recall that the truth vector $w(S_i)$ is the binary vector whose entry corresponds to the term from the full disjunctive normal form (FDNF) of the function $S_i$. FDNF is a disjunctive normal form with disjunctions which contained all variables of the function depends on.

The polynomial expansion of the function is the most efficient representation than others normal forms of Boolean functions for some criteria, e.g. because of a smaller number of terms and units in a circuit (in some cases is much smaller) [6].

As 1 from the truth vector corresponds to the term from FDNF of the function $S_i$, as well as 1 from the Zhegalkin spectrum (or Reed-Muller spectrum [7]) corresponds to the term from the Zhegalkin (Reed-Muller) expansion. This expansion is referred as $r(S_i)$. And the truth vector should be transformed to the Zhegalkin spectrum. This task may be solved with the number of methods, and to demonstrate the procedure of transformation we will use the combinatorial method [8]. The principle of the transformation of $w(S_i)$ to $r(S_i)$ (and backward) for an arbitrary Boolean function $S_i$ is represented with the following theorem.

**Theorem 1 [8].** The $i-$th entry $w_i$ of the truth vector $w(F)=(w_0,w_1,...,w_{2^n-1})$ of the Boolean function $F$ is calculated with the following formula:

$$w_i=\begin{cases}1, if \binom{i}{a_1}+\binom{i}{a_2}+...+\binom{i}{a_q}=1(\text{mod}2);\\ 0-otherwise,\end{cases}$$

where $i=\overline{a_1+1,n}$, $\binom{i}{a_j}=0$ for $i<a_j$ and $w=\Big(\underbrace{0,0,...,0}_{a_1},\underbrace{1,w_{a_1+1},...,w_{2^n-1}}_{n-a_1}\Big)$. In other words, $q$ is the number of unities of the truth vector $w(F)=(w_0,w_1,...,w_{2^n-1})$.

It is helpful to use a consequence of the Lucas theorem [9] to transform $w(S_i)$ to $r(S_i)$.

**Theorem 2 [9].** $\binom{n}{a}=1(\text{mod}2)$ $\Leftrightarrow$ each bit of $a$ is no more than the same bit of $n$.

Let's demonstrate the implementation of theorems on the transformation of $w(S_1)=\{0,1,0,0,1,0,0,1\}$ to $r(S_1)$. According to *Theorem 1* $w_0=r_0=0$ and $w_1=r_1=1$, hence $r(S_1)=(0,1,r_2,...,r_7)$ and using *Theorem 2*

$$r_2=\binom{2}{1}(\text{mod}2)=\binom{10}{01}(\text{mod}2)=0(\text{mod}2)=0,$$

$$r_3=\binom{3}{1}(\text{mod}2)=\binom{11}{01}(\text{mod}2)=1(\text{mod}2)=1,$$

$$r_4=\left(\binom{4}{1}+\binom{4}{4}\right)(\text{mod}2)=\left(\binom{100}{001}+\binom{100}{100}\right)(\text{mod}2)=1(\text{mod}2)=1,$$

$$r_5=\left(\binom{5}{1}+\binom{5}{4}\right)(\text{mod}2)=\left(\binom{101}{001}+\binom{101}{100}\right)(\text{mod}2)=0(\text{mod}2)=0,$$

$$r_6=\left(\binom{6}{1}+\binom{6}{4}\right)(\text{mod}2)=\left(\binom{110}{001}+\binom{110}{100}\right)(\text{mod}2)=1(\text{mod}2)=1,$$

$$r_7=\left(\binom{7}{1}+\binom{7}{4}+\binom{7}{7}\right)(\text{mod}2)=\left(\binom{111}{001}+\binom{111}{100}+\binom{111}{111}\right)(\text{mod}2)=$$
$$=1(\text{mod}2)=1.$$

In the result $r(S_1)=(0,1,0,1,1,0,1,1)$.

As the $q-$ th unity of $r(S_1)=(0,1,0,1,1,0,1,1)$ correspond to the $q-$ th term of the Zhegalin polynomial of the function $S_1$, then $S_1(x_1,x_2,x_3)=x_1\oplus x_1x_2\oplus x_3\oplus x_2x_3\oplus x_1x_2x_3$.

The same procedure is used to generate expansions for $S_2$ and $S_3$.

## III. Software Realization of X (mod P)

The generating of the converter for the calculation of *X (mod P)=S* consists of four steps: calculating of the truth numbers $A(S_i)$ and the truth vector $w(S_i)$ of function $S_i$; transformation of $A(S_i)$ or $w(S_i)$ to $B(S_i)$ and $r(S_i)$ respectively; generating of a polynomial $P(S_i)$; modeling and synthesizing (with ISE Xilinx or LeonardoSpectrum) of the resulting polynomials.

The proposed approach is realized by four software blocks: *Python* $\to$ *Java* $\to$ *Python* $\to$ *VHDL*. The scheme of the software realization in step-by-step manner is pictured at the *Fig*.

Inputs for the first step are values of *P* and *X*. *Python* realization calculates the truth vector $w(S_i)$ and the truth numbers $A(S_i)$. The calculation for $X=(x_{15},x_{14},...,x_1)$ and $P=(p_4,p_3,p_2,p_1)$ is produced in 0,5 second.

The second step is realized by *Java*-block. It transforms of $w(S_i)$ and $A(S_i)$ to $r(S_i)$ and $B(S_i)$ respectively. The process of calculating of $r(S_i)$ and $B(S_i)$ takes approximately 30 seconds for $X=(x_{15},x_{14},...,x_1)$.

The third step is dedicated to generating of all polynomials $P(S_i)$ from $r(S_i)$ ($B(S_i)$), where $i=\overline{1,\delta}$ and $S=(S_\delta,S_{\delta-1},...,S_1)$. The developed *Python* realization produces the step in 10 seconds.

The last fourth step generalizes previous steps. It joins *VHDL* descriptions of all polynomials $P(S_1)$, $P(S_2),\ldots,P(S_\delta)$ in one file. The resulting description is synthesized.

The next section represents the procedure of generating *X (mod P)=S* in details.

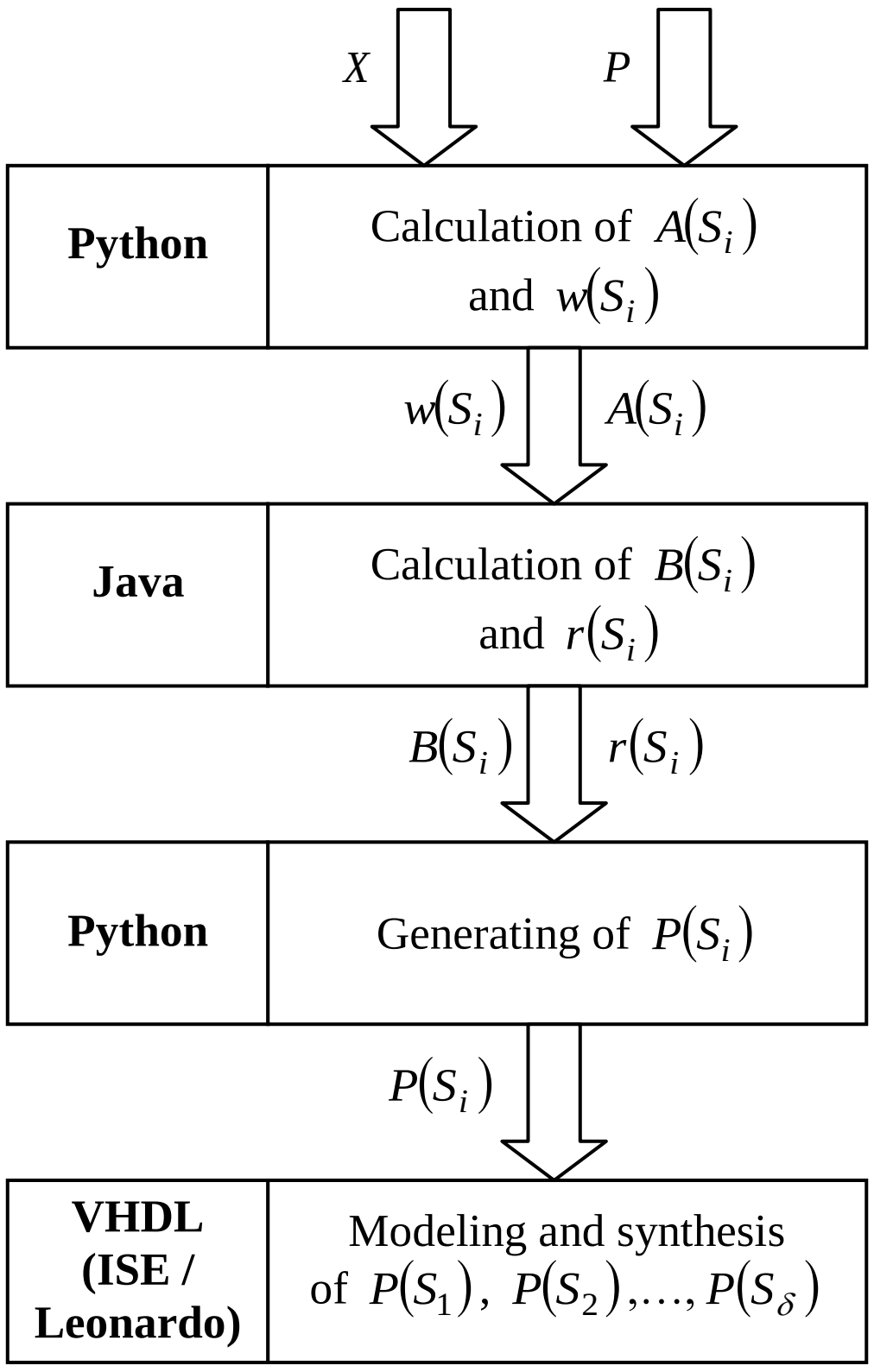

Fig. Structure of the software realization of the proposed approach

## IV. EXAMPLE OF $X\ (MOD\ P)=S$ CALCULATION

Let's consider the process of generating of a converter $X\ (\mathrm{mod} P)=S$ through all steps on the following example, where $X=(x_5,x_4,...,x_1)$ and $P=5$.

### A. *The first step: calculating of* $w(S_i)$

According to the condition $X\in\{0,1,...,31\}$ and $S=(S_3,S_2,S_1)$, sets of the truth numbers for function $S_3,S_2,S_1$ consist of numbers from the range from 0 to 31. The set of the truth numbers and the truth vector for the function:

– $S_1$. The truth numbers with the unity on the 1st bit of numbers for modulo 5 is $A(S_1)=\{1,3,6,8,11,13,16,18,21,23,26,28,31\}$ and the truth vector is $w(S_1)=\{0,1,0,1,0,0,1,0,1,0,0,1,0,1,0,0,1,0,1,0,0,1,0,1,0,0,1,0,1,0,0,1\}$;

– $S_2$. The truth numbers with the unity on the 2nd bit of numbers for modulo 5 is $A(S_2)=\{2,3,7,8,12,13,17,18,22,23,27,28\}$ and the truth vector is $w(S_2)=\{0,0,1,1,0,0,0,1,1,0,0,0,1,1,0,0,0,1,1,0,0,0,1,1,0,0,0,1,1,0,0,0\}$;

– $S_3$. The truth numbers with the unity on the 3rd bit of numbers for modulo 5 is $A(S_3)=\{4,9,14,19,24,29\}$ and the truth vector is $w(S_3)=\{0,0,0,0,1,0,0,0,0,1,0,0,0,0,1,0,0,0,0,1,0,0,0,0,1,0,0,0,0,1,0,0\}$.

### B. *The second step: transformation* $A(S_i)$ *to* $B(S_i)$ *(or* $w(S_i)$ *to* $r(S_i)$*)*

As $A(S_i)$ is analogue to $w(S_i)$, and $B(S_i)$ is analogue to $r(S_i)$, thus we demonstrate this step producing on the transformation of the truth numbers $A(S_i)$ to the Zhegalkin spectrum $r(S_i)$.

Illustration of the transformation is unwieldy. Therefore, we will illustrate by transforming only $A(S_3)$ to $B(S_3)$. From the previous step $A(S_3)=\{4,9,14,19,24,29\}$ and according to *Theorem 1*, and to *Theorem 2* $r(S_3)=(0,0,0,0,1,r_5,...,r_{31})$ and

$r_5=\binom{5}{4}(\mathrm{mod}2)=1$; $r_6=\binom{6}{4}(\mathrm{mod}2)=1$, $r_7=\binom{7}{4}(\mathrm{mod}2)=1$,

$r_8=\binom{8}{4}(\mathrm{mod}2)=0$, $r_9=\left(\binom{9}{4}+\binom{9}{9}\right)(\mathrm{mod}2)=1$,

$r_{10}=\left(\binom{10}{4}+\binom{10}{9}\right)(\mathrm{mod}2)=0$, $r_{11}=\left(\binom{11}{4}+\binom{11}{9}\right)(\mathrm{mod}2)=1$,

$r_{12}=\left(\binom{12}{4}+\binom{12}{9}\right)(\mathrm{mod}2)=1$, $r_{13}=\left(\binom{13}{4}+\binom{13}{9}\right)(\mathrm{mod}2)=0$,

$$r_{14}=\left(\binom{14}{4}+\binom{14}{9}+\binom{14}{14}\right)(\mathrm{mod}2)=0,$$

$$r_{15}=\left(\binom{15}{4}+\binom{15}{9}+\binom{15}{14}\right)(\mathrm{mod}2)=1,$$

$$r_{16}=\left(\binom{16}{4}+\binom{16}{9}+\binom{16}{14}\right)(\mathrm{mod}2)=0,$$

$$r_{17}=\left(\binom{17}{4}+\binom{17}{9}+\binom{17}{14}\right)(\mathrm{mod}2)=0,$$

$$r_{18}=\left(\binom{18}{4}+\binom{18}{9}+\binom{18}{14}\right)(\mathrm{mod}2)=0,$$

$$r_{19}=\left(\binom{19}{4}+\binom{19}{9}+\binom{19}{14}+\binom{19}{19}\right)(\mathrm{mod}2)=1,$$

$$r_{20}=\left(\binom{20}{4}+\binom{20}{9}+\binom{20}{14}+\binom{20}{19}\right)(\mathrm{mod}2)=1,$$

$$r_{21}=\left(\binom{21}{4}+\binom{21}{9}+\binom{21}{14}+\binom{21}{19}\right)(\mathrm{mod}2)=1,$$

$$r_{22}=\left(\binom{22}{4}+\binom{22}{9}+\binom{22}{14}+\binom{22}{19}\right)(\mathrm{mod}2)=1,$$

$$r_{23}=\left(\binom{23}{4}+\binom{23}{9}+\binom{23}{14}+\binom{23}{19}\right)(\mathrm{mod}2)=0,$$

$$r_{24}=\left(\binom{24}{4}+\binom{24}{9}+\binom{24}{14}+\binom{24}{19}+\binom{24}{24}\right)(\mathrm{mod}2)=1,$$

$$r_{25}=\left(\binom{25}{4}+\binom{25}{9}+\binom{25}{14}+\binom{25}{19}+\binom{25}{24}\right)(\mathrm{mod}2)=0,$$

$$r_{26}=\left(\binom{26}{4}+\binom{26}{9}+\binom{26}{14}+\binom{26}{19}+\binom{26}{24}\right)(\mathrm{mod}2)=1$$

$$r_{27}=\left(\binom{27}{4}+\binom{27}{9}+\binom{27}{14}+\binom{27}{19}+\binom{27}{24}\right)(\mathrm{mod}2)=1$$

$$r_{28}=\left(\binom{28}{4}+\binom{28}{9}+\binom{28}{14}+\binom{28}{19}+\binom{28}{24}\right)(\mathrm{mod}2)=0,$$

$$r_{29}=\left(\binom{29}{4}+\binom{29}{9}+\binom{29}{14}+\binom{29}{19}+\binom{29}{24}+\binom{29}{29}\right)(\mathrm{mod}2)=0,$$

$$r_{30}=\left(\binom{30}{4}+\binom{30}{9}+\binom{30}{14}+\binom{30}{19}+\binom{30}{24}+\binom{30}{29}\right)(\mathrm{mod}2)=1,$$

$$r_{31}=\left(\binom{31}{4}+\binom{31}{9}+\binom{31}{14}+\binom{31}{19}+\binom{31}{24}+\binom{31}{29}\right)(\mathrm{mod}2)=0.$$

In the result $r(S_3)=(0,0,0,0,1,1,1,1,1,0,1,0,1,1,0,0,1, 0,0,0,1,1,1,1,0,1,0,1,1,0,0,1,0)$.

In same manner we produce $r(S_1)=(0,1,0,0,0,1,1,1,1,0, 1,0,1,1,0,0,1,0,0,0,1,1,1,1,0,1,0,1,1,0,0,1)$ and $r(S_2)=(0,0,1, 0,0,0,1,1,1,1,0,1,0,1,1,0,0,1,0,0,0,1,1,1,1,0,1,0,1,1,0,0)$.

### C. *The third step: generating of polynomials $P(S_1)$, $P(S_2)$, and $P(S_3)$*

The step aims to generate $P(S_i)$ according to $B(S_i)$ or $r(S_i)$. A unity of the Zhegalkin spectrum indicates which number of term is included in the polynomial. For example, unity on the 6th bit of the Zhegalkin spectrum (it corresponds to the number $6=(x_3x_2x_1)=(110)$ from the set of truth numbers) corresponds to the term $x_2x_3$.

Polynomial expansions for functions $S_1,S_2,S_3$ are represented in VHDL below:

S(1) < = x(1) xor (x(1) and x(3)) xor (x(2) and x(3)) xor (x(1)and x(2) and x(3)) xor x(4) xor (x(2) and x(4)) xor (x(3) and x(4)) xor (x(1) and x(3) and x(4)) xor x(5) xor (x(3) and x(5)) xor (x(1) and x(3) and x(5) ) xor (x(2) and x(3) and x(5)) xor (x(1) and x(2) and x(3) and x(5)) xor (x(1) and x(4) and x(5) ) xor (x(1) and x(2) and x(4) and x(5)) xor (x(3) and x(4) and x(5)) xor (x(1) and x(2) and x(3) and x(4) and x(5));

S(2) <= x(2) xor (x(2) and x(3)) xor (x(1) and x(2) and x(3)) xor x(4) xor (x(1) and x(4)) xor (x(1) and x(2) and x(4) ) xor (x(1) and x(3) and x(4) ) xor (x(2) and x(3) and x(4)) xor (x(1) and x(5)) xor (x(1) and x(3) and x(5)) xor (x(2) and x(3) and x(5)) xor (x(1) and x(2) and x(3) and x(5)) xor (x(4) and x(5)) xor (x(2) and x(4) and x(5)) xor (x(3) and x(4) and x(5)) xor (x(1) and x(3) and x(4) and x(5));

S(3) <= x(3) xor (x(1) and x(3)) xor (x(2) and x(3)) xor (x(1) and x(2) and x(3)) xor (x(1) and x(4)) xor (x(1) and x(2) and x(4)) xor (x(3) and x(4)) xor (x(1) and x(2) and x(3) and x(4)) xor (x(1) and x(2) and x(5)) xor (x(3) and x(5)) xor (x(1) and x(3) and x(5)) xor (x(2) and x(3) and x(5)) xor (x(4) and x(5)) xor (x(2) and x(4) and x(5)) xor (x(1) and x(2) and x(4) and x(5)) xor (x(2) and x(3) and x(4) and x(5)),

where $S_1=S(1)$, $S_2=S(2)$, and $S_3=S(3)$.

### D. *The fourth step: modeling and synthesizing of the resulting polynomials*

This step dedicated to the modeling and synthesis of the VHDL polynomial have been got on the previous step. Synthesis is produced in with a computer-aided-design system (e.g. ISE Xilinx or LeonardoSpectrum).

## V. DISCUSSION AND HARDWARE REALIZATION

This section provides results of comparison of area and the speed of processing between proposed and known approaches. The comparison produced for $X=(x_9,x_8,...,x_1)$ and $P=7$.

Modeling and synthesis is performed in ISE 13.1 and in LeonardoSpectrum2010a_7. The best results in the speed processing (in *ns*) and in the area (in *LUTs*) from ISE and Leonardo were chosen, and they are depictured in two tables.

Table 1 includes results of the synthesis of

– Pipelining approach [1] – *PA*. It is suitable for an arbitrary value of the modulo *P*;

– Iterative approach [2,3] – *IA*. It is suitable for $P=2^n-1$;

– Polynomial expansion approach (proposed approach) – *PEA*. It is suitable for an arbitrary value of the modulo *P*;

– Polynomial expansion approach (proposed approach) after BDD-optimization – *PEA (BDD)*. The optimization [10] of number of terms for the proposed approach was applied. This optimization based on BDD-optimization. It is suitable for an arbitrary value of the modulo *P*.

The synthesis was performed on

– FPGA Xilinx Spartan 3 XC3S1000 FG456 – *Spartan_3*;

– FPGA Xilinx Virtex 7 XC7V285t 3FFG1157 – *Virtex_7*;

– ASIC of the library *POWER* [11], witch is used for design of ASIC circuits on hi-tech factory *Integral* (Minsk, Belarus) – *POWER*, where *UNIT* is an elementary measure of area.

The best indices are highlighted with bold.

TABLE 3. COMPARISON OF RESULTS OF THE SYNTHESIS

| | | FPGA | | ASIC |
|---|---|---|---|---|
| | | Spartan_3 | Virtex_7 | POWER |
| PA | T, ns | 35,078 | 13,828 | 35,67 |
| | LUTs / UNITs | 73 | 34 | 88 393 |
| LA | T, ns | 12,033 | 6,599 | 9,7 |
| | LUTs / UNITs | **14** | **10** | **23 313** |
| PEA | T, ns | 12,585 | **6,45** | 14,96 |
| | LUTs / UNITs | 68 | 33 | 307 815 |
| PEA (BDD) | T, ns | **11,532** | 7,547 | **6,91** |
| | LUTs / UNITs | 58 | 31 | 37 124 |

Table 2 demonstrates speed characteristics (in *ns*) and in area units for FPGAs (in *LUTs*) for some prime modules. The synthesis performed for 10-bits range *X* and for 3-, 4, and 5-bits modulo *P*. As well the Table 2 consists best results of the synthesis in ISE 13.1 and in LeonardoSpectrum2010a_7 of the proposed polynomial expansion approach – *PEA*.

TABLE 3. RESULTS OF THE SYNTHESIS OF *X (MOD P)* FOR X [10:1]

| | P | 7 | 11 | 13 | 17 | 19 | 23 | 29 | 31 |
|---|---|---|---|---|---|---|---|---|---|
| Spartan 3 | time, ns | 12,725 | 14,308 | 14,753 | 12,847 | 19,708 | 16,207 | 16,774 | 13,232 |
| | LUTs | 92 | 179 | 171 | 140 | 245 | 255 | 303 | 98 |
| Virtex 7 | time, ns | 8,28 | 9,397 | 9,429 | 9,43 | 9,266 | 9,522 | 9,827 | 9,122 |
| | LUTs | 49 | 81 | 105 | 114 | 126 | 136 | 161 | 100 |

Table 3 contains an example of the synthesis for *P=691* and it is a prime number. The synthesis executed using the proposed approach and with the conditions as for the above examples.

TABLE 3. RESULTS OF THE SYNTHESIS OF *X (MOD 691)* FOR X [11:1]

| | P | 691 |
|---|---|---|
| Spartan 3 | time, ns | 13,912 |
| | LUTs | 128 |
| Virtex 7 | time, ns | 8,789 |
| | LUTs | 111 |

## VI. CONCLUSIONS AND FURTHER WORK

There are two main advantages of the proposed approach for *X (mod P)* realization: flexibility of *P*, because *P* can be an arbitrary number, and no memory hardware realization, because it is used only XOR and AND operators.

The *Table 1* provides the results of comparison for the case $P=2^3-1$. As we see, the proposed technique has the speed processing advantage over the known realizations.

Theoretically the proposed approach of hardware realization of *X (mod P)* goals to calculate the operation for an arbitrary *X* and *P*. But the bottleneck is in the realization *X (mod P)* for a big range of *X*, e.g. if *X [30:1]* and *P=7* the realization process takes more than 24 hours. The process of the synthesis on FPGA takes most of this time.

In this way, the proposed approach has two directions for improving: area optimization and expanding the range of *P*.

Primarily further work will be directed to getting as short as possible polynomials and, as a consequence, reducing of the hardware complexity of the scheme of converter. The progress could be achieved at the expense of expanding of the range of *X*.